# Pressure sensing using vertically aligned carbon nanotubes on a flexible substrate

*E. L. Carter, P. Brown, R. L. Smith and J. Griffin*

*Sullivan University, Louisville, KY, USA*

*Abstract:*

*Sensing technologies have been under research and development for their varied applications from microelectronics to space exploration. With the end of Moore's law in sight, there is growing demand for shrinking materials and improving sensitivity and range of sensing of sensors. Carbon nanotubes (CNTs) offer an excellent combination of small size (in the order of nanometers in two dimensions and micrometers in the third dimension), varied current conductivity (from insulating to metallic), flexibility, mechanical strength and feasibility of mass production. Here we used CNTs to fabricate pressure sensors to sense static loads of pressure and studied the characteristics of different methods of building the sensors. We offer an adhesive-absorption technique of fabrication of pressure sensors that tackles the issue of endurance of the sensors to repeated operation. We demonstrate a significant change in resistance of a vertically aligned forest of nanotubes upon application of static loads. This study will enable building of better pressure sensors for several applications.*

The extreme efforts in advancing Moore's law is accompanied by demands to discover new materials that can perform better functions in terms of computation and responsive systems. One important such system is pressure sensor. Carbon nanotubes (CNTs) have paved way for realizing nanotechnology at the system level due to their early discovery relative to other nanomaterials. This has led to several research efforts in building sensing systems with CNTs. There are many problems that need to be addressed before CNTs become a routinely used material for pressure sensing. Among the issues faced are: low endurance to repeated operation of the pressure sensor, lack of sensing capabilities, uncertainty on techniques to grow dense CNTs and issues related to signal processing of the measured electrical signal. Here we grew dense

CNT forests of vertically aligned CNTs and transferred them onto a flexible substrate, namely kapton. We show that the pressure sensing capabilities of the CNT forest is consistent and excellent in terms of endurance, signal levels and simplicity of construction.

We grew a dense forest of vertically aligned CNTs using Al+Fe thin films as catalysts, as described in our earlier work. Following this, we used our earlier described technique of transferring the forest of CNTs onto flexible substrates. Kapton was the best choice for adhesion and stiction. Using a high temperature processing, we annealed the transferred films at 600 K for 2 hours to improve adhesion. This was followed by vacuum baking to remove moisture. The setup to sense pressure is displayed in Figure 1. The pressure is applied on the top of the vertically aligned CNT forest. The electrical resistance of the forest is read using two leads connecting the two sides of the forest. About 1 cm x 1cm area was chosen to perform the experiments. Pressure was applied using a variety of static weights and a disc approximately the size equal to the area of the patch of CNTs to equally distribute the pressure across the entire area. We expected the CNTs to bend and hence produce a change in resistance, which could be read by the electrical leads from either sides of the system. Thus we measured only the in-plane resistance of the forest of CNTs and as we will show later, this was an effective way of sensing the highest possible signal from this setup.

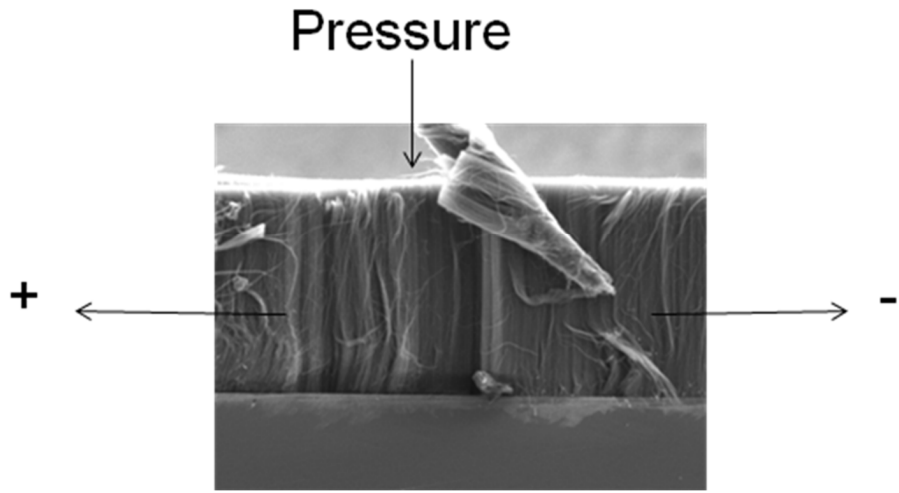

**Figure 1:** SEM image of the CNT forest; height of the CNTs being 0.16 mm. Also shown is the configuration for pressure sensor using the CNT layer. '+' and '-' indicate the points across which the resistance was read.

The first configuration involved testing of the pressure sensing capabilities of the CNT forest on top of the silicon wafer on which it was originally grown (without any transfer). For applied pressure, the response of the resistance of the CNT forest is displayed in Figure 2. The pressure indicated here is measured in arbitrary units in view of the fact that the area of the CNT patch can be arbitrarily define and the size of the pressure disc can also be arbitrarily defined. It is seen that as pressure increases, the resistance of the film decreases. This can be intuitively understood by considering the additional crowding effect faced by the CNT forest upon application of a pressure from the top of the CNT forest. As pressure is applied, the CNTs are bound to bend and near the position of the bend, the neighboring CNTs are forced closer to one another. This creates a more continuous electrical conduction pathway, thereby reducing the resistance of the films. The ratio of change in resistance is approximately 2. While this is not very high, it was very repeatable.

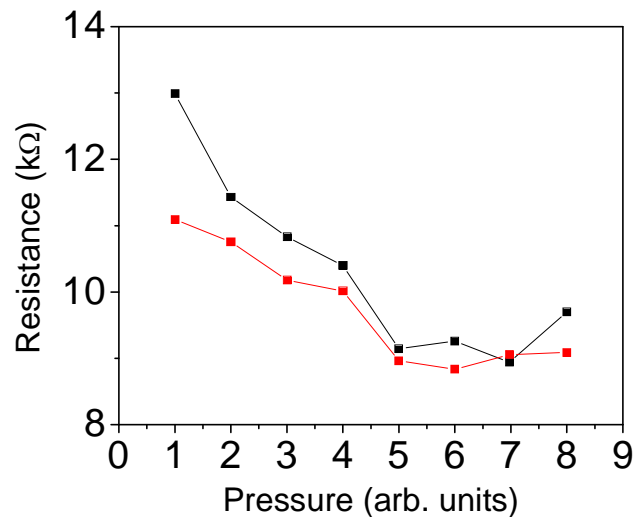

**Figure 2:** Response of CNTs on Si substrate to pressure.

The second configuration involved transfer of CNTs without using an adhesive onto kapton substrate. The rest of the transfer process remained identical to the one described in the previous paragraph. For applied pressure, the response of the resistance of the CNT forest is displayed in Figure 3. As before, the pressure is defined by arbitrary units for similar reasons described above. It is seen that, similar to the previous case of pressure sensing using a forest of CNTs on silicon wafer, the electrical resistance decreases as pressure is increased. The ratio of resistance change is much worse compared to the case of CNT forest on silicon wafer.

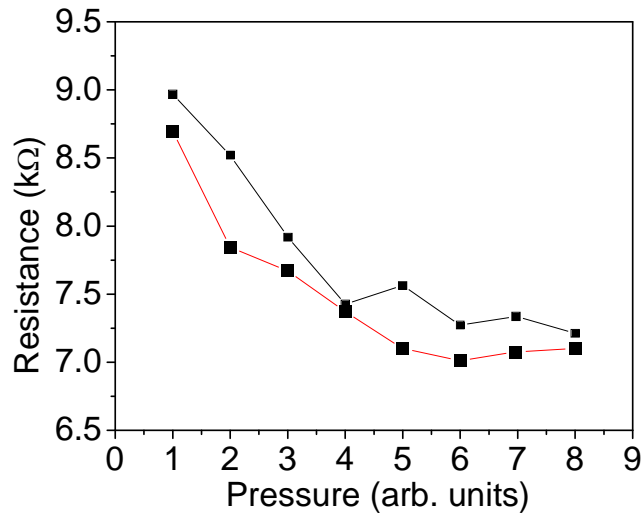

**Figure 3:** Response of CNTs on kapton substrate without adhesive to pressure.

One of the important issues with the pressure sensing was repeatability of the measurement. The first time pressure is applied, the electrical response to the pressure is significant, while the subsequent times, it is not significant. In order to correct this issue, we used an adhesive absorption technique, where we pour adhesive on top of the CNT forest such that the adhesive will get absorbed into the CNT forest and also get dried inside the forest. Id the adhesive is chosen in such a way that the dried form has significant elastic limits and elasticity, the forest can 'spring back' every time it is pressurized. Figure 4 is a depiction of this process in schematics. Upon complete absorption of the adhesive into the CNT forest, the advantages of this approach are: (1) The original alignment of the CNTs are retained upon repeated pressurizing, (2) no migration of the CNT forest occurs, (3) No aggregation or agglomeration of the CNT forest occurs, which is a common cause of failure of CNT forests and CNT solutions and (4) no voids are present in the CNT forest, which are filled by the adhesive.

Following the addition of the adhesive, we tested a forest of CNTs for pressure sensing. The results are displayed in Figure 5. The change in resistance followed a similar trend compared to the ones seen before, but the ratio of the change in resistance was much higher (about 6). This increase in ratio of resistance was accompanied by increased endurance to repeated cycles of application of pressure.

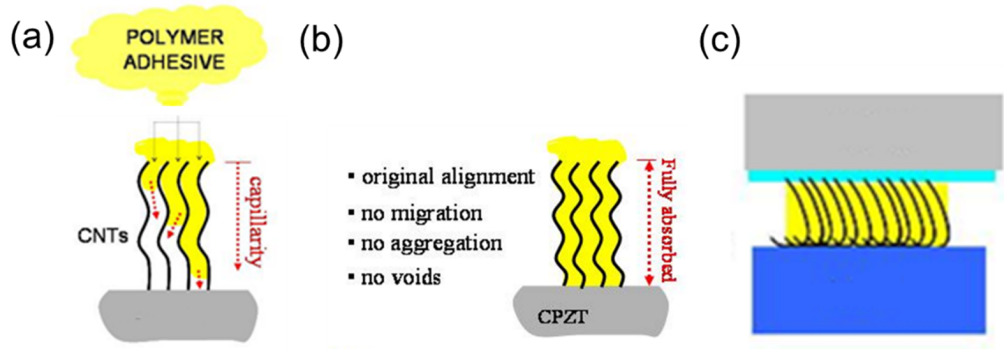

**Figure 4:** (a) Schematic of adhesive addition to the CNT forest. (b) Schematic of complete adhesive absorption into the CNT forest. (c) Schematic of bending of CNTs with applied pressure.

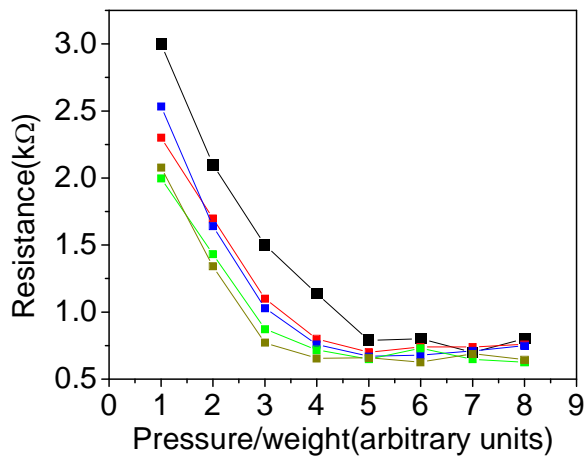

**Figure 5:** Response of CNTs on kapton substrate with adhesive to pressure.

Following 100 cycles of application of pressure to the CNT forest with adhesive, we imaged the CNT forests in a scanning electron microscope to look for features indicating fatigue/damage. Figure 6 displays the results of this study. Figure 6a is a SEM image of the CNT forest with adhesive after 100 cycles. This shows very little buckling following the pressure application. Figure 6b is a SEM image of the CNT forest without adhesive after 100 cycles. There is clear buckling of the forest. This is evidence that explains why the CNT forest without adhesive did not offer a high endurance.

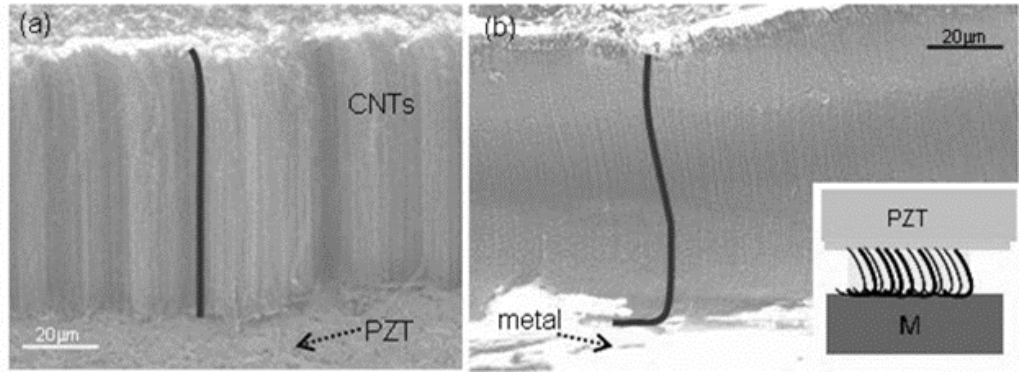

**Figure 6:** (a) CNT forest with adhesives after 100 cycles of application of pressure. (b) CNT forest without adhesives after 100 cycles of application of pressure.

As a part of future work, The results with adhesive addition are promising. We could further improve the adhesion of the CNTs to the Kapton substrate by using higher temperatures and pressures. Since this method has not been tried before, we have a lot of optimization that can be done. The transfer process could also involve a patterned metal layer, which forms the patters of the required geometry, for any given application. In this case, the nanotubes forests are isolated in smaller islands and are connected by the underlying metal itself. This is more like a self assembled circuit, where the active part of the device and interconnects are grown/printed in the same process. For better testing, we could use a precision low force testing setup, since we do not have one right now. This could be as simple as an AFM or an Omniprobe. While measuring the resistance across the length of the CNT layer is done here, it could also be done across the thickness of the layers. This will require another couple of metallization steps.

In conclusion, we demonstrated that we can grow CNT forests on silicon wafers and transfer them onto flexible substrates like kapton. This was followed by a demonstration of pressure sensing capability of the forests. We showed that addition of adhesives to the forest can greatly improve the endurance of the pressure sensing and also the signal levels produced by the experiment. We showed using direct evidence that the buckling of the CNT forests under pressure is highly curtailed by the addition of the adhesive. This process is likely to aid in large scale production of pressure sensors that will last long.